\documentclass[a4paper,10pt]{article}

\usepackage{graphicx,float}

\usepackage[cp1251]{inputenc}

\usepackage{cite,amsmath,amsfonts,amsthm,fullpage}
\usepackage{youngtab}

\usepackage{graphics}
\usepackage{miniplot}
\usepackage{subfigure}

\newcommand{\pb}{\mathbf{p}}

\theoremstyle{plain}

\newtheorem{Lemma}{Lemma}
\newtheorem{Proposition}{Proposition}
\newtheorem{Corollary}{Corollary}
\newtheorem{Remark}{Remark}
\newtheorem{Example}{Example}

\theoremstyle{remark}

\def\tr{\mathrm {tr}}
\def\det{\mathrm {det}}

\def\bp{\begin{Proposition}}
\def\ep{\end{Proposition}}
\def\bc{\begin{Corollary}}
\def\ec{\end{Corollary}}
\def\bl{\begin{Lemma}}
\def\el{\end{Lemma}}
\def\be{\begin{equation}}
\def\ee{\end{equation}}
\def\br{\begin{Remark}\rm\small}
\def\er{\end{Remark}}
\def\brs{\begin{remarks}.\\ \rm\
\begin{enumerate}}
\def\ers{\end{enumerate}\end{remarks}}
\def\bea{\begin{eqnarray}}
\def\eea{\end{eqnarray}}

\def\bx{\begin{Example}\rm\small}
\def\ex{\end{Example}}

\def\tr{\mathrm {tr}}
\def\det{\mathrm {det}}

\def\&{&{\hskip -20pt}}

\newcount\YDcount\YDcount=0
\def\YDsize{10pt}

\def\YD#1{%
\ifnum#1=0
 \ifnum\YDcount=0 \ifx\varnothing\undefined\emptyset\else\varnothing\fi
 \else\vskip1.4pt\egroup\YDcount=0\fi
\else
 \ifnum\YDcount=0 \YDcount=1\vcenter\bgroup\vskip1pt
 \else\nointerlineskip\fi
 \vbox{\hrule\hbox{\vrule height\YDsize
 \loop\hskip\YDsize\vrule\ifnum\YDcount<#1\advance\YDcount1\repeat}\hrule
 \kern-0.4pt}\expandafter\YD
\fi}

\usepackage[usenames,dvipsnames]{color}
\usepackage{ulem}
\usepackage{graphicx}

\begin{document}

\author{ Aleksander Yu.
Orlov\thanks{Institute of Oceanology, Nahimovskii Prospekt 36,
Kurchatov Institue,
Moscow, Russia, email: orlovs@ocean.ru }}
\title{Polygon gluing and commuting bosonic operators}

\date{June 30, 2020}
\maketitle

To Leonid Chekhov and Nikita Slavnov in connection with their 60th anniversary

\begin{abstract}
Two families of commuting Hamiltonians are constructed, parametrized by a constant matrix. The first series was guessed and is new, while the second is known, and in our approach it follows from the first series. For proof, I used the facts known from our previous consideration of the relationships between random matrices and Hurwitz numbers, but the text is selfconsistent and does not require reference to these works.
\end{abstract}

\bigskip


$\,$

\section{Introduction}

The problem of finding a ring of commuting operators acting in an infinite-dimensional space is one of the problems of mathematics and mathematical physics. We found an example of such a ring. \footnote{I was initiated by the articles \cite{MM1}-\cite{MirMor} which are researches on a slightly different topic.}
Modern methods for obtaining such rings are described.
in various works, in particular, these are works on quantum integrable systems. We can't even look through the huge collection of articles on these topics. To search for links, I present only
articles \cite{MolevNazOlshansli},\cite{Zheglov}
with a wide list of references, and I must be sorry for the incomplete bibliography. Here I present a family of operators tested for commutation among themselves. The proof is simple, but I use some facts that we know very well from the relationship between Ginibre multimatrix ensembles and polygon gluing, see \cite{NO2020},\cite{NO2020F}. In fact, the steps of the proof are quite elementary. The text is consistent and it's just a straight calculation with a little trick to make it easier.

\section{Commuting Hamiltonins I}

Consider the following oscillator algebra:
\be
[\phi^\dag_{i,j},\phi^\dag_{i',j'}]=
[\phi_{i,j},\phi_{i',j'}]=0,\quad 0\le i,j,i',j'\le N
\ee
\be
[\phi^\dag_{i,j},\phi_{j',i'}]=\delta_{i,i'}\delta_{j,j'},\quad 0\le i,j,i',j'\le N
\ee
The Fock space is generated by the action of the creation operators $\phi_{i,j},\,0\le i,j \le N$ on the vacuum vector $|0\rangle$, and
\be
\phi^\dag_{i,j}|0\rangle =0,\quad 0\le i,j \le N.
\ee

Matrices with entries $\phi_{i,j},\,0\le i,j \le N$
and $\phi^\dag_{i,j},\,0\le i,j \le N$ we denote $\phi$
and $\phi^\dag$ respectively.

Introduce
\be
H_n(A) =\tr\left((\phi^\dag \phi A)^n \right)
\ee

If $A$ is a Hermitian matrix, then the operators $H_n(A)$ can be regarded as the Hamiltonians of the system of bosonic particles. It can be noted that for $A=I_N$ the operator $H_2(I_N)$ can be associated with the Laplace-Beltrami operator on the group $GL_N$ and with the Hamiltonian of the quantum Calogero system at the special value of the coupling constant, when the eigenfunctions of the Hamiltonian turn into Schur functions:
\be
H_n(I_N)s_\lambda(\phi C)|0\rangle =E_n(\lambda)s_\lambda(\phi C)|0\rangle,\quad C\in GL_N
\ee
where $\lambda$ is a partition, $E_n(\lambda)$ is an eigenvalue, and $C$ is independent of $\phi$.
\br

From the consideration in \cite{NO2020F} it follows:
\be\label{shift}
H_n(A)s_\lambda(\phi C)|0\rangle =E_n(\lambda)s_\lambda(\phi AC)|0\rangle,
\ee
where $|\lambda|=n$.
The relation (\ref{shift}) can be interpreted as an eigenvalue problem in two cases (i) either $A=I_N$ or (ii) $A\neq I_N,\,AC=C$, which implies that both $ A$ and $C$ degenerate. Case (i) is a specialization of the "generalized cut and join equation" introduced in \cite{MMN2}, and for this case (\ref{shift}) is true without the restriction $|\lambda|=n$.

\er

A remarkable property of the operators $H_n(A)$ is the fact that they commute with each other:
\bp \label{proposition1}

We get
\be
[H_n,H_m]=0
\ee
for any pair of $n,m\in \mathbb{Z}_\ge$.
\ep
 Proof.
 Consider the multicomponent analogue of the operators 
 $\phi$ and $\phi^\dag$:
 \be
[\phi^{(a)\dag}_{i,j},\phi^{(b)}\dag_{i',j'}]=
[\phi^{(a)}_{i,j},\phi^{(b)}_{i',j'}]=0,\quad 0\le i,j,i',j'\le N
\ee
\be
[\phi^{(a)\dag}_{i,j},\phi^{(b)}_{j',i'}]=\delta_{i,i'}\delta_{j,j'},\quad 0\le i,j,i',j'\le N
\ee
\be
\phi^{(a)\dag}_{i,j}|0\rangle =0,\quad 0\le i,j \le N.
\ee
for $a,b=1,\dots,k$.

 Let's represent the creation operator $\phi^{(a)}_{i,j}$ as an arrow, the beginning of which will be labeled $i$ and the end labeled $j$. Let's draw it as a white dotted arrow, and assign the number $a$ to the arrow itself. The annihilation
 operator $\phi^{(a)\dag}_{i,j}$ will be represented by a black dotted arrow with the same number $a$ , but now we assign the label $j$ to the beginning of the operator this arrow and the label $i$ to end.
  
 Then each entry $B^{(a)}_{i,j}$ we depict by a solid black arrow with the number $a$, we assign the label $i$ to its beginning, and we attribute the label $j$ to the end.
Similarly, each entry $C^{(a)}_{i,j}$ is depicted as a solid white arrow with the number $a$, we attribute the label $i$ to its beginning, and the label $j$ to the end.
  
 The trace of the product of several matrices can be represented by a polygon
   with directed edges with summation of labels at the vertices. We have two such products: the first one comes from $H_n(A)$ and the second one comes from $H_m(A)$ Which we call black and white polygons resectively.  The edges of the black polygon will be directed in the positive direction (counterclockwise) and the edges of the white polygon will be directed negatevely.
   That is, the transition counterclockwise from one black polygon edge to another edge corresponds to reading the product under the trace sign from left to right and 
   the transition clockwise from one white polygon edge to another edge also corresponds to reading the product under the trace sign from left to right. If we draw a negatively oriented k-gon on a sphere, then its complement is a positively oriented k-gon. In what follows, gluing $k$-gons will be understood as gluing a black $k$-gon with a complement to a white $k$-gon, see Fig. 1 with triangles for the visualization.

    We consider two oppositely oriented $2k$-gonons: the black one generated by $\tr \left(\phi^{(1)\dag} B_1 \phi^{(2)\dag} B_2\cdots
        \phi^{(k)\dag} B_k \right)$ whose edges are alternating dotted and solid black arrows,
        and white one generated by $\tr \left(\phi^{(k)} C_k \phi^{(k-1)} C_{k-1}\cdots
        \phi^{(1)} C_1 \right)$ whose edges are alternating dotted and solid white arrows.  
      
 In fact, it is more convenient to first consider the case when all matrices $B_i,C_i,\,i=1,\dots,k$ are identical ones. In this case, all edges of both polygons are shown as dotted arrows, see the see the top images in the Fig.1. And only then add to this drawing the matrices $B_i,C_i,\,i=1,\dots,k$ at the corners of the polygons as crossbars (solid arrows) at the corners, see the second from the top of the image in the Fig.1. Therefore, we will call the matrices $B_i$ and $C_i$ the black and white corner matrices, respectively.
 
 One can varify
  \be\label{Sphere}
  \langle 0|\tr \left(\phi^{(1)\dag} B_1 \phi^{(2)\dag} B_2\cdots 
  \phi^{(k)\dag} B_k \right)
  \tr \left(\phi^{(k)} C_k \phi^{(k-1)} C_{k-1}\cdots 
  \phi^{(1)} C_1 \right)|0\rangle =
  \tr (B_1C_2)\tr(B_2C_3)\cdots \tr(B_kC_1)
  \ee
  This relation can be interpreted as a gluing of two
    $2k$-gons as follows.
    Each pair  of oppositely directed dotted arrows (one black, one white) with number $a$ ($a=1,\dots,k$) are glued according to the rule
  \[
   \langle 0| \phi^{(a)\dag_{i,j}}\phi^{(b)}_{j',i'}
   |0\rangle = \delta_{a,b}\delta_{i,i'}\delta_{j,j'}
  \]
 One can see that (\ref{Sphere}) describes the sphere glued from two polygons with $k$ vertices each is surrounded by negatively oriented 2-gones whose edges are given by the arrows related to the matrices $ (B_1,C_2),(B_2,C_3)\cdots,(B_k,C_1)$.
 Then, the right hand side in (\ref{Sphere}) is obvious.
 
 It can be seen that (\ref{Sphere}) describes a sphere glued from two polygons with $k$ vertices, each of which is surrounded by negatively oriented 2-gons, the edges of which are given by arrows associated with the matrices $(B_1,C_2), (B_2 ,C_3)\cdots,(B_k,C_1)$. (This sphere can be thought of as a globe glued together from two hemispheres, on the equator of which there are $k$ vertices, each of which is surrounded by a digon, two edges of which belong to the northern and southern hemispheres, respectively).
   Then the right side in (\ref{Sphere}) is obvious.
  
 However, which surface is obtained from the mathematical expectation depends on which edges of the white polygon with which edges of the black polygon
  we glue. If we renumber the edges
   white polygon like
   $(k,k-1,\dots,1)\to i_k,i_{k-1},\dots,i_1$ and glue the polygons again along the dotted arrows with the same numbers, then another oriented surface can be obtained.
 
 For instance, see the top part of Fig.1. We have
 \be\label{sphere'}
 \langle 0|\tr\left(\phi^{(1)\dag}B_1\phi^{(2)\dag}B_2\phi^{(3)\dag}B_3\right)
 \tr\left( \phi^{(2)}C_2\phi^{(1)}C_1\phi^{(3)}C_3 \right) |0\rangle = \tr (B_1C_2) \tr(B_2C_3) \tr(B_3C_1)
 \ee 
  since it turns out a sphere obtained from two triangles
   (two hemispheres: southern (black) and northern (white)).
   Here are the faces of the black triangle
    are numbered $1,2,3$ if we go around it in a positive direction relative to the North Pole. The edges of the white triangle are numbered in the same way if we go around it in a positive direction relative to the South Pole. We get 3 vertices, each of which has two corners and monodromy of each  the vertex is the product of one black and one white corner matrix.
  
 The expression
  \be\label{torus}
 \langle 0|\tr\left(\phi^{(1)\dag}B_1\phi^{(2)\dag}B_2\phi^{(3)\dag}B_3\right)
 \tr\left( \phi^{(1)}C_1\phi^{(2)}C_2\phi^{(3)}C_3 \right) |0\rangle = \tr(B_1C_2B_3C_1B_2C_3)
 \ee
 describes the gluing of two triangles with edges numbered $1,2,3$ and $2,1,3$, resulting in a torus with one 6-valent vertex and 6 neighboring corners of alternating colors,
 see the bottom half of the figure 1. 

 \begin{figure}[h]
 \includegraphics[scale=0.8]{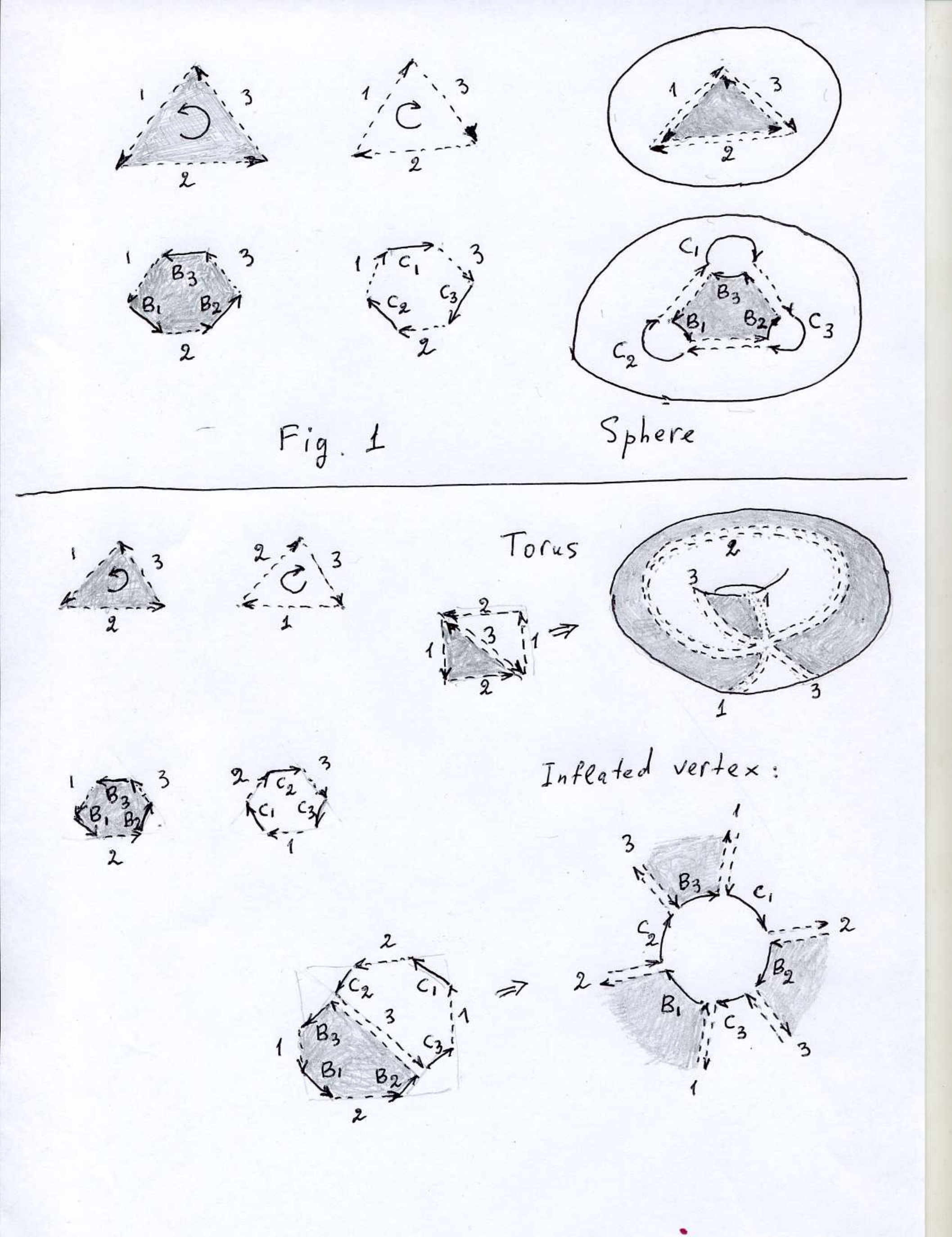}
\caption{Gluing black and white triangles}
\label{MyGreatPicture1}
 \end{figure}

 We get 
 \bl\label{general'}
   \be\label{general}
  \langle 0|\tr \left(\phi^{(1)\dag} B_1 \phi^{(2)\dag} B_2\cdots 
  \phi^{(k)\dag} B_k \right)
  \tr \left(\phi^{(i_k)} C_{i_k} \cdots \phi^{i_{1}}
  C_{i_k} \right)|0\rangle =
  \tr (W_1)\cdots \tr(W_V)
  \ee
 where $V$ is the number of vertices, and each $W_\alpha$ is a monodromy obtained as the product of corner matrices around the vertex $\alpha$ while traversing it clockwise.
 \el
 \br \label{alternating}
 An important statement is that the corner matrices associated with each vertex alternate between black and white as you go around the vertex. The vertex $\alpha$ of valence $2v$ has a monodromy of the form $W_\alpha =(B_{i_1}C_{j_1})\cdots (B_{i_v}C_{j_v})$, where the numbers of the corner matrices $i_1,\dots,i_v$ and 
 $j_1,\dots,j_v$ belong to the set $1,\dots,k$. For example, see the right hand sides of the equations (\ref{Sphere}), (\ref{sphere'}), (\ref{torus}), see figure 1 for a pair of examples.
 \er
 
Now, one can write
\be\label{BC}
:\tr \left( \left( \phi^\dag \phi A\right)^n\right) :
\, 
:\tr\left(\left(\phi^\dag \phi A\right)^m \right) :=
\sum_{k=0}^{\min (n,m)} :Q_k:
\ee
  where $k$ is the number of couplings 
  $\langle 0|\phi^\dag \phi |0\rangle$ according to the Wick rule. 
Each $:Q_k:$ is the sum which includes (i) the sum over the choice of samples of $k$ operators $\phi^\dag$ in the first multiplier  in (\ref{BC}) and the sum over the choice of samples $\phi$ in the second multiplier  in the left hand side of (\ref{BC}) (ii) the sum over the choice of pairing in the chosen samples which is the sum over elements of the permutation group $S_k$. 

where $k$ is the number of pairings
    $\langle 0|\phi^\dag \phi |0\rangle$ according to Wick's rule.
Each $:Q_k:$ is the sum including (i) the sum over the choice of samples from $k$ $\phi^\dag$ operators in the first factor in (\ref{BC}) and the sum over the choice of $\phi$ samples in the second factor on the left hand side (\ref{BC}) (ii) the sum over the choice of pairing in the selected samples, which is the sum over the elements of the permutation group $S_k$.

\br\label{numbering}
To describe different pairings we number the chosen $\phi^\dag$ with the numbers $1,\dots,k$ in
\be\label{n}
\underbrace{(\phi^\dag\phi A)\cdots (\phi^\dag\phi A)}_n
\ee
in the order from the left to the right. In the product
\be\label{m}
\underbrace{(\phi^\dag\phi A)\cdots (\phi^\dag\phi A)}_m
\ee
we number the chosen $\phi$ with the numbers $\sigma(1),\dots,\sigma(k)$ from the left to the right. We couple the $\phi^\dag$ to $\phi$ with the same number. 

One should have in mind that under the sign of the trace
the product a cyclic permutation does cange the trace. Then we can place the of the $\phi^\dag_1$ and $\phi_1$ on the leftmost place resepctively in (\ref{n}) and 
(\ref{m}).
\er

Symbolically one can write
\be\label{samplesBC}
  :Q_k:=\sum_{\rm samples}\sum_{\sigma\in S_k} :q_\sigma(k\times k\,{\rm samples}):
\ee
  
  The terms $:Q_0:,:Q_1:$ and $:Q_2:$ are the easiest ones: 
  $$
  :Q_0:=:\tr \left( \left( \phi^\dag \phi A\right)^n\right) \, 
\tr\left(\left(\phi^\dag \phi A\right)^m \right) :
$$ 
$$
  :Q_1:= nm :\tr \left( A\left( \phi^\dag \phi A\right)^{n+m-1}\right)  :
$$
$$
  :Q_2:= nm\sum_{n_1,n_2,m_1,m_2\atop n_1+n_2=n-2,\,m_1+m_2=m-2}
  :\tr \left( A\left( \phi^\dag \phi A\right)^{n_1+m_2-1}\right) 
 \tr \left( A\left( \phi^\dag \phi A\right)^{n_2+m_1-1}\right)  :
$$
For $k=3$ we use (\ref{sphere'}) and (\ref{torus}):
$$
  :Q_3:= 
$$
$$
  nm\sum_{n_1,n_2,n_3,  
  \, n_1+n_2+n_3=n-3 \atop  m_1,m_2,m_3,\,
  m_1+m_2+m_3=m-3}
  :\tr \left( A\left( \phi^\dag \phi A\right)^{n_1+m_2-1}\right) 
 \tr \left( A\left( \phi^\dag \phi A\right)^{n_2+m_3-1}\right)  
 \tr \left( A\left( \phi^\dag \phi A\right)^{n_3+m_1-1}\right):
$$
$$
+ nm\sum_{n_1,n_2,n_3,  
  \, n_1+n_2+n_3=n-3 \atop  m_1,m_2,m_3,\,
  m_1+m_2+m_3=m-3}
 : \tr\left(  A\left( \phi^\dag \phi A\right)^{n_1+m_2-1}
   A\left( \phi^\dag \phi A\right)^{n_3+m_1-1}
   A\left( \phi^\dag \phi A\right)^{n_2+m_3-1}\right)
  :
$$
  
  For each $k$ we have
  \be
  :Q_k:=n\sum'_{n_1,\dots,n_k\atop m_1,\dots,m_k}\sum_{\sigma\in S_k} :q_\sigma(k):
  \ee
  where $\sigma$ enumerates different ways of couplings 
  of each $\phi$ among chosen $k$ samples from the left
  hand side to the chosen $k$ samples of $\phi^\dag$ 
  from the left hand side. There are $ \left(n\atop k \right)$ and $\left(m\atop k \right)$ ways to choose the samples and there are $|S_k|=k!$ different pairings (in fact $(k-1)!$ different pairings because of the cyclic permutations). Actually these numbers are not important for us.

When the choice of samples and the numbering are fixed one can write the left side of (\ref{BC}) in form (\ref{general}) where
\be\label{B_iC_i}
B_i=\phi A\left(\phi^\dag \phi A \right)^{n_i},\quad
C_i=A\left(\phi^\dag \phi A \right)^{m_i}\phi^\dag
\ee
where the numbers $n_1,\dots,n_k$ measure a 'distance' between neiboring $\phi^\dag$ chosen in (\ref{n}). There are the following equalities
\be\label{conditions_n}
\sum_{a=1}^v n_{i_a}=n-k
\ee
\be\label{conditions_m}
\sum_{a=1}^v m_{j_a}=m-k
\ee

  Each way of pairing yields a certain embedded graph. Each graph gives rise to a
  product of traces of the monodromies around the vertices, see Lemma \ref{general'} where the structure of each monodromy is clarified by the Remark \ref{alternating}, therefore the trace of each monodromy $W_\alpha$
  has a form
  \be
 \tr W_\alpha= \tr\left( (B_{i_1}C_{j_1})\cdots(B_{i_v}C_{j_v}) \right)=\tr \left(A(\phi^\dag\phi A)^{n_{i_1}+m_{j_1}+1}\cdots A (\phi^\dag\phi A)^{n_{i_v}+m_{j_v}+1}   \right)
  \ee
  with some $i_1,j_1,\dots,i_v,j_v$
  where $2v$ is the valence of a vertex $\alpha$.
Thus each trace of monodrome is completely characterized 
by the ordered set 
\be\label{spect}
n_{i_1}+m_{j_1},\dots,n_{i_v}+m_{j_v}
\ee
given up to a cyclic permutation. Let us call it the spectrum of the vertex $\alpha$.

Now we will compare these spectra.

 Similarly for the second term in the commutator we write 
 \be\label{C^*B^*}
:\tr\left(\left(\phi^\dag \phi A\right)^m \right) :
:\tr \left( \left( \phi^\dag \phi A\right)^n\right) :
\, =
\sum_{k=0}^{\min (n,m)} :Q^*_k:
\ee
and
\be\label{samplesC^*B^*}
  :Q^*_k:=\sum_{{\rm samples}^*}\sum_{\sigma\in S_k} :q^*_\sigma(k\times k\,{\rm samples}^*):
\ee
We will compare terms $:q_\sigma(k\times k\,{\rm samples}):$ and $:q^*_\sigma(k\times k\,{\rm samples}^*):$
for a given $\sigma$.
We also relates the samples as follows:
to each chosen $\phi^\dag$ from a sample in (\ref{samplesBC}) we choose the nearest right neiboring 
$\phi$ for the sample in (\ref{samplesC^*B^*}) and to each chosen $\phi$ of a sample in (\ref{samplesBC}) we
choose the nearest left neiboring $\phi^\dag$ for the associated sample in (\ref{samplesC^*B^*}). Such pairs of samples we will call dual ones.

  where $k$ is the number of pairings 
  and for each $k$ we have
  \be
  :Q^*_k:=\sum_{\sigma\in S_k} :q^*_\sigma(k):
  \ee

 \be
 \tr W^*_\alpha= \tr\left( (B_{i_1}C_{j_1})\cdots(B_{i_v}C_{j_v}) \right)=\tr \left(A(\phi^\dag\phi A)^{n_{i_1}+m_{j_2}+1}\cdots A (\phi^\dag\phi A)^{n_{i_v}+m_{j_1}+1}   \right)
  \ee 
which is characterized by the spectrum 
\be\label{spect^*}
n_{i_1}+m_{j_2},\dots,n_{i_v}+m_1
\ee
Now one compare this dual spectrum with the spectrum 
(\ref{spect}). Spectrums (\ref{spect}) and (\ref{spect^*}) can be one-to one related by
the cyclic permutation of the numbers $n_{i_a},\,a=1,\dots,v$ related to the vertex $\alpha$ under consideration. Namely, by
\be
n_{i_1}\to n_{i_2},\dots, n_{i_v}\to n_{i_1}
\ee
and all $m_{j_a},\,a=1,\dots,v$ in (\ref{spect^*}) are
the same as in (\ref{spect}). These replacements do
not violate conditions (\ref{conditions_n}) and 
(\ref{conditions_m}).

The same correspondences between terms resulting from (\ref{BC}) and from (\ref{C^*B^*}) we get for all vertices of the resulting graph, and one gets it for all graphs. The proof is complete.

For a given partition $\mu=(\mu_1,\dots,\mu_\ell$, $\ell=1,2,\dots$ we introduce
\be
H_\mu(A)=
:\prod_{i=1}^\ell \tr\left(\left(\phi^\dag \phi A\right)^{\mu_i}\right) :
\ee

In the same way as we prove Proposition \ref{proposition1} one can prove that
\be
[H_n(A)\,,\, 
H_\mu(A)]=0
\ee
for $n=1,2,\dots$ and any partition $\mu=(\mu_1,\mu_2,\dots)$.

In this case for each given $k$ we glue a single black $k$-gon related to the picking out the discribed samples in $\tr\left(\left(\phi^\dag \phi A\right)^n\right)$ to the
set of white polygones obtained as picking out of the bosonic operators from the product $h_\mu(A)$. In this case we use the same cyclic permutation of the numbers $n_i$ described above.

\section{Commuting Hamiltonins II}

Suppose $A=1+\epsilon \textsc{a}$, where 
$\textsc{a}\in {\rm Mat}_{N\times N}$, $\epsilon$
is a parameter.

Let us introduce
\be
h_n(A)=:\tr\left(a\left(\phi^\dag\phi\right)^n  \right)
\ee
and
\be
h_\lambda(A)=:\prod_{i=1}^\ell \tr\left(a\left(\phi^\dag\phi\right)^{\lambda_i}  \right):
\ee
where $\lambda=(\lambda_1,\dots,\lambda_\ell$ is a partition.
Then from Proposition \ref{proposition1} and from 
\bl\label{center}
\be
\left[H_n(I_N),H_m(I_N)\right] = 0,
\ee
from
\el
\bl\label{center_Gurevich}
\be
\left[H_n(I_N),h_\lambda(a)\right] = 0
 \ee
\el
it follows
\begin{Corollary} Let $\textsc{a}\in {\rm Mat}_{N\times N}$ 
 \be\label{Gurevich}
\left[h_n(a),h_\lambda(a)\right] = 0
 \ee
\end{Corollary}

Each of Lemmas (\ref{center}),(\ref{center_Gurevich}) and actually (\ref{Gurevich}) is proven in the same way as Proposition 1. We omit the details.

Being written in the form of differential operators and without the sign of normal ordering 
Lemmas (\ref{center}),(\ref{center_Gurevich}) and also (\ref{Gurevich})
are quite known, for example see 
\cite{SaponovGurevich}.

\section{Discussion}

(i) We found a series of commuting Hermitian operators,
this means that these is a quantum integrable
system. 
We mark that the eigenfunctions for the Hamiltonians 
$h_n(A)$ are yet unknown in case of general Hermitian 
matrices. This is a problem to be solved together
with a number of problems, say, solved in \cite{Slavnov}
for a number of quantum integrable problems.

(ii) It will be interesting to compare these results to the results obtained in \cite{Olshanski-19},\cite{Olshanski-199},\cite{Okounkov-199},\cite{Okounkov-1996},\cite{MolevNazOlshansli},\cite{NazOlsh},\cite{Nazarov-Sklyanin2009},\cite{Sharygin},\cite{Sharygin1},\cite{Gurevich},\cite{SaponovGurevich}.

(iii) Links with the quantum Calogero model is the other interesting direction to work out. In this relation the following works should be pointed out \cite{MM1}-\cite{MirMor},\cite{Resh},\cite{SergeevVeselov2013},\cite{Gur}

\section*{Acknowledgements}

The author is grateful to George Shurygin for important discussions and to A.D.Mironov, D.Gurevich and  A.Zheglov for inspirating talks on the related subjects. I also thank A.E.Mironov and I.A.Taimanov for the invitations to Novosibirsk where this work was finalized.
The work was supported by the Russian Science
Foundation (Grant No.20-12-00195).

\appendix

\section{Partitions and Schur functions \label{Partitions-and-Schur-functions-}}

Let us recall that the characters of the unitary group $\mathbb{U}(N)$ are labeled by partitions
and coincide with the so-called Schur functions \cite{Mac}. 
A partition 
$\lambda=(\lambda_1,\dots,\lambda_n)$ is a set of nonnegative integers $\lambda_i$ which are called
parts of $\lambda$ and which are ordered as $\lambda_i \ge \lambda_{i+1}$. 
The number of non-vanishing parts of $\lambda$ is called the length of the partition $\lambda$, and will be denoted by
 $\ell(\lambda)$. The number $|\lambda|=\sum_i \lambda_i$ is called the weight of $\lambda$. The set of all
 partitions will be denoted by $\mathbb{P}$.

The Schur function labelled by $\lambda$ may be defined as  the following function in variables
$x=(x_1,\dots,x_N)$ :
\be\label{Schur-x}
 s_\lambda(x)=\frac{\det \left[x_j^{\lambda_i-i+N}\right]_{i,j}}{\det \left[x_j^{-i+N}\right]_{i,j}}
 \ee
 in case $\ell(\lambda)\le N$ and vanishes otherwise. One can see that $s_\lambda(x)$ is a symmetric homogeneous 
 polynomial of degree $|\lambda|$ in the variables $x_1,\dots,x_N$, and $\deg x_i=1,\,i=1,\dots,N$.
  
 \br\label{notation} In case the set $x$ is the set of eigenvalues of a matrix $X$, we also write $s_\lambda(X)$ instead
 of $s_\lambda(x)$.
 \er

 There is a different definition of the Schur function as quasi-homogeneous non-symmetric polynomial of degree $|\lambda|$ in 
 other variables, the so-called power sums,
 $\pb =(p_1,p_2,\dots)$, where $\deg p_m = m$.
 
For this purpose let us introduce 
$$
 s_{\{h\}}(\mathbf p)=\det[s_{(H_i+j-N)}(\mathbf p)]_{i,j},
$$
where $\{h\}$ is any set of $N$ integers, and where
the Schur functions $s_{(i)}$ are defined by $e^{\sum_{m>0}\frac 1m p_m z^m}=\sum_{m\ge 0} s_{(i)}(\pb) z^i$.
If we put $H_i=\lambda_i-i+N$, where $N$
is not less than the length of the partition $\lambda$, then
\begin{equation}\label{Schur-t}
 s_\lambda(\mathbf p)= s_{\{h\}}(\mathbf p).
\end{equation}

 The Schur functions defined by (\ref{Schur-x}) and by (\ref{Schur-t}) are equal,  $s_\lambda(\pb)=s_\lambda(x)$, 
 provided the variables $\pb$ and $x$ are related by the power sums relation
  \be
\label{t_m}
  p_m=  \sum_i x_i^m
  \ee
  
  In case the argument of $s_\lambda$ is written as a non-capital fat letter  the definition (\ref{Schur-t}),
  and we imply the definition (\ref{Schur-x}) in case the argument is not fat and non-capital letter, and
  in case the argument is capital letter which denotes a matrix, then it implies the definition (\ref{Schur-x}) with $x=(x_1,\dots,x_N)$ being
  the eigenvalues.

 Relation (\ref{Schur-t}) relates polynomials $s_\lambda$
 and $\pb_\Delta$ of the same degree $d=|\lambda|=|\Delta|$.
 Explicitly one can write
\be\label{powersum-Schur}
\pb_{\Delta}=\sum_{\lambda\in\Upsilon_d} 
\frac{{\rm dim}\lambda}{d!}
\zeta_\Delta\varphi_\lambda(\Delta) s_\lambda(\pb)
\ee
 and
\be\label{Schur-char-map}
s_\lambda(\pb)=\frac{{\rm dim}\lambda}{d!}\sum_{\Delta\in\Upsilon_d } 
\varphi_\lambda(\Delta)\pb_{\Delta}.
\ee
The last relation is called the character map relation.
Here 
\be\label{dim}
 \frac{{\rm dim}\lambda}{d!}: =
\frac{\prod_{i<j\le N}^{}(\lambda_i-\lambda_j-i+j)  }{\prod_{i=1}^{N}(\lambda_i-i+N)!}
\ee
(see example 1 in sect 1 and example 5 in sect 3 of chapt I in \cite{Mac}), where   $N\ge \ell(\lambda)$. As one can check, the right hand side does not depend on $N$.
(We recall that $\lambda_i=0$ in case $i>\ell(\lambda)$. The number
${\rm dim}\lambda$ is an integer.

The factors $\varphi_\lambda(\Delta)$ satisfy the following orthogonality relations
\be\label{orth1}
\zeta_{\Delta}\sum_{\lambda \in\Upsilon_d} 
\left(\frac{{\rm\dim}\lambda}{d!}\right)^2\varphi_\lambda(\mu)\varphi_\lambda(\Delta) =
 \delta_{\Delta,\mu} 
\ee
and
\be\label{orth2}
\left(\frac{{\rm\dim}\lambda}{d!}\right)^2
\sum_{\Delta\in\Upsilon_d} \zeta_\Delta\varphi_\lambda(\Delta)\varphi_\mu(\Delta) =
\delta_{\lambda,\mu}.
\ee

\section{Operators $H_\lambda(A)$ and Hurwitz numbers}

The consideration of multimatrix models (like considered in \cite{Chekhov}) and
Hurwitz numbers in \cite{NO2020},\cite{NO2020F} 
can be naturaly reformulated as the relations of the bosonic Hamiltonians $H_\lambda(A)$ and Hurwitz numbers ${\cal H}()$ which enumerate coverings of Riemann sphere with three  branch points yields 
\be\label{Hur-bos}
H_\lambda(A)\prod_{i} \tr\left(\left(\phi C \right)^{\mu_i}\right)|0\rangle =
\sum_{\nu_i} {\cal{H}}\left(\lambda,\mu,\nu \right)
\tr\left(\left(\phi AC \right)^{\nu_i}\right)|0\rangle
\ee
where $\lambda,\mu,\nu$ are partitions conditioned by
$|\lambda|=|\mu|=|\nu|$ which define ramification profiles in  the branch points and ${\cal{H}}\left(\lambda,\mu,\nu \right)$ is the Hurwitz number.
It has the meaning of the structural constant in a Frobenius algebra related to Hurwitz numbers.

By the characteristic map relations (\ref{Schur-char-map}) and by (\ref{orth1}),(\ref{orth2}) it can be rewritten in form
\be\label{MMN-A-bos}
H_\lambda(A) s_\mu\left(\left(\phi C \right)\right)|0\rangle =
\frac{|\mu|!}{\dim \mu}\chi_\mu(\lambda)
s_\mu\left((\phi AC \right)|0\rangle
\ee
where $\chi_\mu(\lambda)$ is the character of the irreducable representation $\mu$ of the symmetric group $S_d,\,d=|\mu|$ evaluated on the cycle class $\lambda$.
In case $A=C=I_N$ it is a particular form of
Mironov-Morozov-Natanzon cute-and-join greneralized relation \cite{MM1}. Related problems were discussed in  \cite{PerelomovPopov2}, see also \cite{Zhelobenko} for some links.

\end{document}